\documentclass{article}

\PassOptionsToPackage{numbers, compress}{natbib}

\usepackage[final]{neurips_2024}

\usepackage[utf8]{inputenc} 
\usepackage[T1]{fontenc}    
\usepackage{hyperref}       
\usepackage{url}            
\usepackage{booktabs}       
\usepackage{amsfonts}       
\usepackage{nicefrac}       
\usepackage{microtype}      
\usepackage{xcolor}         

\newcommand{\sparagraph}[1]{\vspace{1mm}\noindent {\bf #1}}
\newcommand{\BtrBlocks}{\textsc{BtrBlocks}}
\newcommand{\FastLanes}{\textsc{FastLanes}}
\newcommand{\ri}{(\texttt{i})}
\newcommand{\rii}{(\texttt{ii})}
\newcommand{\framework}{\texttt{Virtual}}
\newcommand{\totalamount}{\texttt{total\_amount}}

\usepackage{float}
\usepackage{enumitem}
\usepackage{amsmath}
\usepackage{bm}
\usepackage{graphicx}
\usepackage{subcaption}
\usepackage{caption}
\usepackage{float}
\usepackage{amssymb}
\usepackage{wrapfig}
\usepackage{colortbl}
\usepackage{siunitx}
\usepackage{amsthm}

\definecolor{customgreen}{rgb}{0.0, 0.5, 0.0}
\definecolor{customred}{rgb}{0.8, 0.0, 0.0}

\usepackage{multirow}
\usepackage{booktabs}

\DeclareMathOperator*{\argmin}{arg\,min}

\makeatletter
\newcommand{\subalign}[1]{%
  \vcenter{%
    \Let@ \restore@math@cr \default@tag
    \baselineskip\fontdimen10 \scriptfont\tw@
    \advance\baselineskip\fontdimen12 \scriptfont\tw@
    \lineskip\thr@@\fontdimen8 \scriptfont\thr@@
    \lineskiplimit\lineskip
    \ialign{\hfil$\m@th\scriptstyle##$&$\m@th\scriptstyle{}##$\hfil\crcr
      #1\crcr
    }%
  }%
}
\makeatother

\title{Lightweight Correlation-Aware Table Compression}

\author{%
  Mihail Stoian, Alexander van Renen, Jan Kobiolka, Ping-Lin Kuo,\\
  \textbf{Josif Grabocka, Andreas Kipf}\\
  \texttt{\{mihail.stoian,~alexander.van.renen,~jan.kobiolka,~ping-lin.kuo,}\\\texttt{josif.grabocka,~andreas.kipf\}@utn.de} \\
  University of Technology Nuremberg
}

\begin{document}

\maketitle


\begin{abstract}
    The growing adoption of data lakes for managing relational data necessitates efficient, open storage formats that provide high scan performance and competitive compression ratios. While existing formats achieve fast scans through lightweight encoding techniques, they have reached a plateau in terms of minimizing storage footprint. Recently, correlation-aware compression schemes have been shown to reduce file sizes further. Yet, current approaches either incur significant scan overheads or require manual specification of correlations, limiting their practicability. We present \texttt{Virtual}, a framework that integrates seamlessly with existing open formats to automatically leverage data correlations, achieving substantial compression gains while having minimal scan performance overhead. Experiments on \texttt{data.gov} datasets show that \texttt{Virtual} reduces file sizes by up to 40\% compared to Apache Parquet.
\end{abstract}

\section{Introduction}

When a comet passes the Earth, mankind carefully prepares well in advance; months, if not years.
We see something similar with the constant shift to the cloud: Large relational tables are moving to the cloud, and we are not there yet in terms of efficient processing times. This is partly because current storage formats, such as Apache Parquet, cannot keep up with the ever-increasing data sizes. Simply put, scan performance and compression ratios are still poor. In fact, Parquet itself applies general-purpose compression, e.g., Snappy, on top to improve its storage footprint further. As a result, latest research, e.g., \BtrBlocks{}~\cite{btr_blocks}, \FastLanes{}~\cite{fast_lanes}, tries to adapt to the new setting by ensuring two main requirements: \ri{} column scans are fast, \rii{} compression ratios remain competitive.

While requirement \ri{} has been pushed to its limits, the second still awaits its solution. The key insight is that existing techniques use only simple encoding schemes~\cite{lemire, lightweight_survey, cost_based_lightweight, chimp} which naturally leads to a plateau in terms of minimizing storage footprint.

\sparagraph{State of the Art.}~For this very reason, recent research proposes a change in mindset: Instead of relying only on simple encoding schemes such as \texttt{RLE} or \texttt{DELTA}, leverage correlations between columns that are present in real-world datasets~\cite{deepsqueeze, white_box, cuttle, ticc, corbit, corra, c3_glas} and represent them in a compact way; the standard encoding schemes can be applied on top. This is known as semantic or correlation-aware compression and has been shown to enable \rii{}. However, existing solutions for numeric columns either do not satisfy \ri{} or are not user-friendly: \textsc{DeepSqueeze}~\cite{deepsqueeze} is based on an auto-encoder architecture that naturally makes a column dependent on all the other columns; this incurs a slowdown during column scans and violates~\ri{}. \textsc{Corra}~\cite{corra}, while lightweight in itself, relies on user-defined functions to actually be able to compress. \textsc{C3}~\cite{c3_glas} automates this process, but only applies a simple linear regression to column pairs for its numeric columns, leaving compression potential unexplored. \textsc{Cuttle}~\cite{cuttle} and \textsc{TICC}~\cite{ticc} target more general functions defined on column-tuples, but they do not optimize for the number of reference columns that have to be loaded during scans.



\sparagraph{Contribution.}~We introduce \framework{}, a lightweight framework that \emph{automatically} detects \emph{sparse} functions inherently present in a given table, thus enabling both (\texttt{i}) fast scan times and (\texttt{ii}) competitive compression ratios.~\framework{} is designed to be a plug-in for open storage formats. In particular, we show saving factors of up to 40\% over Parquet on \texttt{data.gov} tables and a minimal slowdown of column scans on the virtualized columns.~\framework{} is open source and available at \url{https://github.com/utndatasystems/virtual}.

\section{Lightweight Correlation-Aware Compression}

\framework{} has three main components: (\texttt{a}) \textsc{FunctionDriller} (Sec.~\ref{sec:driller}), (\texttt{b}) \textsc{Optimizer} (Sec.~\ref{sec:optimizer_and_compressor}), and (\texttt{c}) \textsc{Compressor} (Sec.~\ref{sec:optimizer_and_compressor}). This corresponds to three natural steps: (\texttt{a}) Find the inherent functions in the table, (\texttt{b}) choose which functions to select from the previously found set, and then (\texttt{c}) perform the actual compression using the functions. In the following, we describe (\texttt{a}) in more detail.


\subsection{\textsc{FunctionDriller}}\label{sec:driller}

Our function ``driller'' determines which tuples of columns are correlated and the associated functions.

\sparagraph{(Simple) Example.}~As an introductory example, consider the real-world dataset ``\emph{Monthly Modal Time Series}'' within \texttt{data.gov}: It has a total of thirteen column names prefixed with ``\emph{Total}''. Not surprisingly, all these columns are the sum of specific columns in the dataset -- we refer to these as \emph{reference} columns. The to-be-virtualized columns are referred to as \emph{target} columns. Yet, no current open storage format exploits this correlation. As we show in this work, considering these column correlations results in high saving rates.

\sparagraph{(Complex) Example.}~Not all correlations are as explicit as the previous one. Prior work has shown that the well-studied Taxi~\cite{taxi_2018} table has multiple functions underlying its \totalamount{} column~\cite{corra}. This is the motivation behind what will be introduced next as $K$-regression. Note that \citet{corra} identified the different functions by hand, while \framework{} finds them automatically.

\sparagraph{Drilling via $K$-Regression.}~To find the above functions in a principled manner, we fix a target column $y \subset T$ from a given table $T$ and find an estimator using the remaining columns, $X := T \setminus y$. Specifically, we use a regression model known in the literature as \emph{cluster-wise} linear regression~\cite{clusterwise1, clusterwise2}. The task is to partition the table records into $K$ clusters such that the sum of squared errors of the linear regressions across clusters is minimized. Let $\left(C_1, C_2, \dots, C_K \right)$ represent the cluster partitions, where $C_k$ represents the set of row indices assigned to the $k$-th cluster. Furthermore, let $ \boldsymbol{w}_k, \beta_k$ be the weight vector and the intercept for the linear model of cluster $C_k$. 

We initialize the model by randomly assigning each data row $i$ to one of $K$ distinct clusters and subsequently estimate a linear regression model for each cluster. In the \textbf{first step}, we assign each row $i$ to the  optimal cluster $k^{(i)}$ computed as the cluster whose linear model best approximates the target column $y_i$ of the table row $i$: 
\begin{equation}
    \forall i: \;\; C_{k^{(i)}} \leftarrow C_{k^{(i)}} \cup \{i\}, \text{ where   } k^{(i)} = \argmin_{k \in \left\{1,\dots,K\right\}} \left( y_i - X_i^\top \boldsymbol{w}_k - \beta_k \right)^2.
    \label{eq:clusterassignment}
\end{equation}
Given the cluster assignments, in the \textbf{second step}, we then optimize the linear regression models as:
\begin{equation}
    \argmin_{\bm{w}, \beta}  \sum_{k=1}^{K} \sum_{i \in C_k} \left( y_i - X_i^\top \bm{w}_k - \beta_k \right)^2.
    \label{eq:k_regression}
\end{equation}
The method iterates the two steps of Eq.~\eqref{eq:clusterassignment} and Eq.~\eqref{eq:k_regression} until convergence (for practical reasons, only on a randomly selected sample of $10^4$ rows). However, using this naively does not satisfy one of our main requirements: ``\emph{scans should be fast}''. To see why this is the case, note that a plain linear regressor is not optimized for a small number of reference columns. We alleviate this issue with a sparsity regularization approach.

\sparagraph{Why Sparseness Matters.}~Correlation-aware compression, by design, introduces overhead for subsequent table scans, which may lead to impractical downstream data analyses. To understand this limitation, consider the following SQL query (a), which computes the average of the total amounts across all taxi rides~\cite{taxi_2018}:
\begin{figure}[H]
\centering 
\begin{enumerate}[label=(\alph*)]
    \item \texttt{select avg(total\_amount) from taxi;} \hfill \textcolor{customgreen}{\texttt{2.7 ms}}\\
    \item \texttt{select avg(fare\_amount + mta\_tax + tip\_amount) from taxi;} \hfill \textcolor{customred}{$\uparrow$ \texttt{6.5 ms}}
\end{enumerate}
\end{figure}
Naturally, if \totalamount{} is now calculated by the sum of other columns as in SQL query (b), the execution engine\footnote{In this case, we used DuckDB~\cite{duckdb} and ingested the first 10M rows of the table.} needs to load these columns to evaluate the expression. As the example shows, this leads to extra I/O which slows down the query by a linear factor depending on the number of reference columns. Hence, to minimize scan overhead, we should aim for the \emph{minimum} number of reference columns that have to be loaded.

\begin{figure*}
    \centering
    \includegraphics[width=1.0\linewidth]{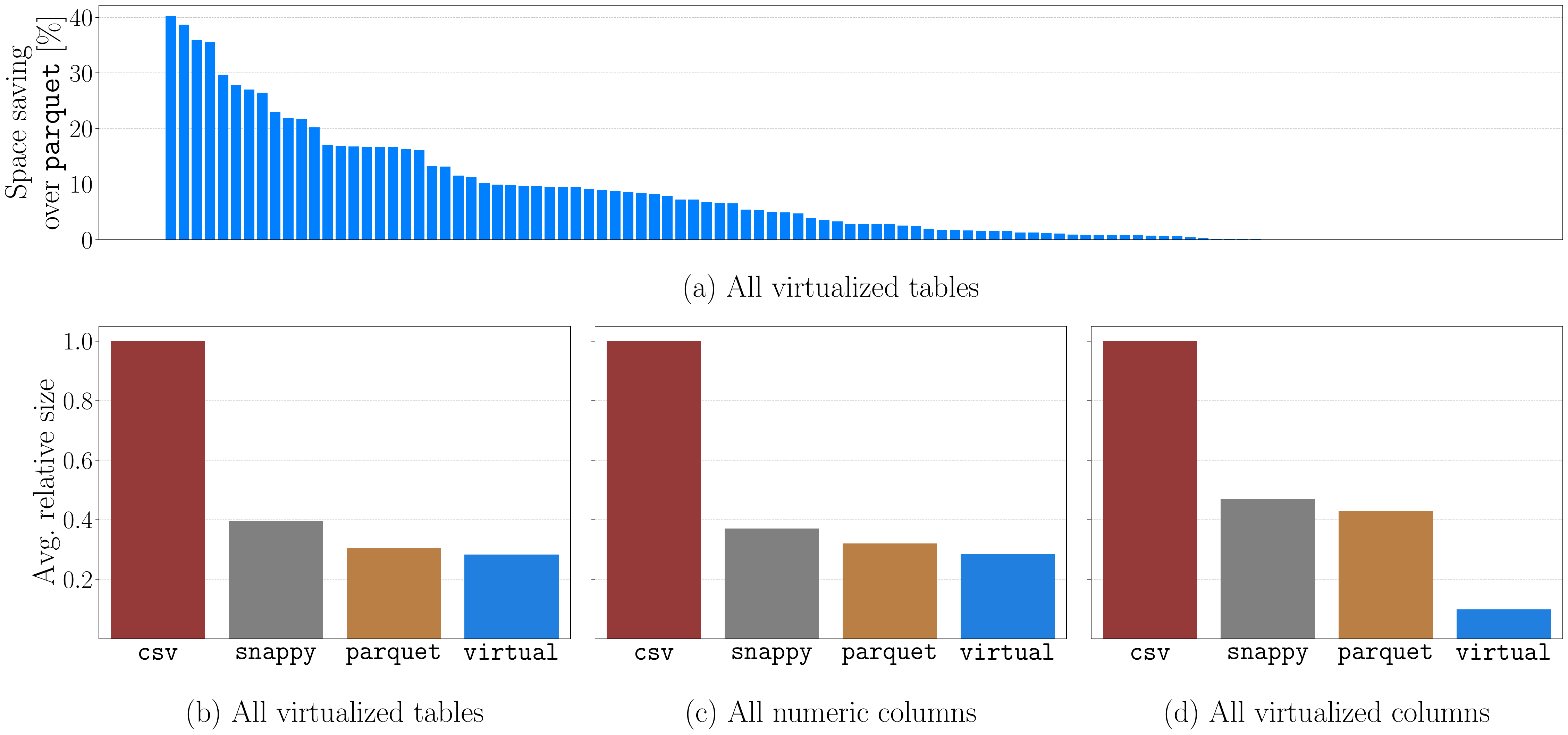}
    \caption{Improvement over Parquet+Snappy (\texttt{parquet}) across 103 virtualized \texttt{data.gov} tables}
    \label{fig:main_plot}
\end{figure*}

Formally, this leads to a \emph{sparse} linear regression~\cite{sparse_reg}, i.e., the number of non-zero entries of the estimator should be minimized using a $L_0$ penalty. Hence, instead of minimizing Eq.~\eqref{eq:k_regression}, we have:
\begin{equation}
    \argmin_{\bm{w}, \beta}  \sum_{k=1}^{K} \sum_{i \in C_k} \left(y_i - X_i^\top \bm{w}_k - \beta_k \right)^2 + \lambda \|\bm{w}\|_0,
    \label{eq:sparse}
\end{equation}
where $\|\bm{w}\|_0$ is defined as the number of non-zero coefficients of the coefficient vector $\bm{w}$; notably, there is constraint for the intercept $\beta$ since it does not involve the loading of any reference column.





\subsection{\textsc{Optimizer} \& \textsc{Compresser}}\label{sec:optimizer_and_compressor}

\sparagraph{Conflicting Functions.}~Once the $K$-regressor has been found, we can start virtualizing the columns. Yet, we are still missing an important aspect: Not all columns can actually be virtualized. To see why this is the case, consider the following example: Column ``$\texttt{a}$'' can be computed from ``$\texttt{b} + \texttt{c}$''. However, this also holds for ``$\texttt{b}$'' and ``$\texttt{c}$'', e.g., ``$\texttt{b} = \texttt{a} - \texttt{c}$''. Indeed, virtualizing any column-pair would lead to a cyclic dependency between the columns.~We use an optimizer that estimates the space saving of the found functions and greedily chooses a valid subset to maximize the estimated gain.



\sparagraph{Compression.}~The selected target columns output by the optimizer can now be virtualized. The auxiliary columns to store are as follows: (a) \texttt{offset} — which stores the offset to the \emph{closest} of the $K$ regressors, and (b) \texttt{switch} — which indicates in $\log_2 K$ bits which of the $K$ regressors has been chosen.\footnote{In the case of $K = 1$, this auxiliary column is not considered.} However, when applying correlation-aware compression to real-world tables, the missing value issue becomes more stringent. Intuitively, they add another layer of complexity, as evaluating expressions on \texttt{NaN} values results in such a value. Unlike prior work, we choose not to drop these rows but instead also store the following auxiliary columns: (c) \texttt{outlier} — in case one of the reference values is missing (\texttt{NaN}), and (d) \texttt{is\_nan} — which indicates whether the original value was itself a \texttt{NaN}. This layout ensures that data can be reconstructed in a lossless manner.

\section{Experiments}

We perform a large-scale evaluation to understand the applicability of \framework{} to real-world tables. We run on a \texttt{c5d.4xlarge} instance with 16 vCPUs, 32 GB of memory, and 500 GB of disk space. We downloaded the most popular CSV tables from \texttt{data.gov}~\cite{data_gov}, obtaining 1,226 datasets.\footnote{A number of 150 datasets had downloading issues, showing a pending update; we already excluded these.} As relational tables are usually stored in data blocks, we restrict ourselves to the first $10^6$ rows of each table. Moreover, note that we guarantee a \emph{lossless} compression, as is common in databases.

\sparagraph{Compression Ratios.}~Of the 1,226 tables, we were able to find correlations in 157 of them (we selected regressors with error less than $1$). We also dropped tables with less than 1,000 rows, since, in these scenarios, the Parquet files were larger than the original CSV files. We are left with 103 tables, whose space savings over Parquet+Snappy we show sorted in Fig.~\ref{fig:main_plot}(a). The largest space saving of 40\% is for a geo-location dataset~\cite{geo}. In Fig.~\ref{fig:main_plot}(b-d), we show relative sizes to CSV, with Fig.~\ref{fig:main_plot}(d) best expressing the effect of virtualization on a total of 328 columns.
\begin{wrapfigure}{r}{0.5\textwidth}
    \includegraphics[width=0.45\textwidth]{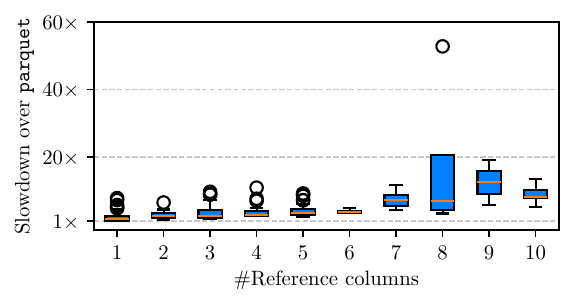}
    \caption{Column scan slowdowns}
    \label{fig:plot_slowdown}
    \vspace{-0.6cm}
\end{wrapfigure}

\sparagraph{Scan Times.}~We enable fast column scans for correlation-aware compression by relying on sparse $K$-regressors.~In Fig.~\ref{fig:plot_slowdown}, we show the slowdown over Parquet+Snappy when summing up the values of virtualized columns directly from the files generated by default Parquet and \framework{}, respectively.\footnote{Note that for robust measurements, we duplicate the rows until each table reaches 1M rows.} Since the auxiliary columns (Sec.~\ref{sec:optimizer_and_compressor}) are lightweight (the \texttt{offset} is minimized via Eq.~\eqref{eq:k_regression}), the overhead scales linearly with the number of reference columns.




\sparagraph{Comparison to \texttt{DeepSqueeze}~\cite{deepsqueeze}.}~We also compare to \texttt{DeepSqueeze}, an auto-encoder architecture that provides excellent compression ratios. We use its publicly available re-implementation~\cite{DeepSqueeze_github} (note that this only supports numeric columns;~\citet{deepsqueeze} do not specify how many distinct values make up a categorical column).~Since \texttt{DeepSqueeze} is lossy by design, we made the following changes: We (\texttt{1}) removed input quantization -- essentially the ``lossy'' component in \texttt{DeepSqueeze}, (\texttt{2}) added offsets post-rescaling, (\texttt{3}) rounded the results to the correct precision, and (\texttt{4}) stored rows with \texttt{NaN}s in a separate Parquet file. Regarding~(\texttt{2}), we found that adding offsets after rescaling empirically reduced the file size for integer columns (compared to storing small floating-point offsets). In Tab.~\ref{tab:comparison}, we show the file size and column scan latency for three tables from \texttt{data.gov}. In particular, \texttt{DeepSqueeze} becomes impractical in terms of column scans as it has to load the \emph{entire} latent space to perform a single column scan.

\begin{table}[ht]
\centering
\begin{tabular}{c@{\quad}S[table-format=2.2]S[table-format=2.2]@{\qquad}S[table-format=4.2]S[table-format=2.2]}
  \toprule
  \multirow{2}{*}{\shortstack{Metric\\~\\~\\Approach}} & \multicolumn{2}{c}{File size [MB]} & \multicolumn{2}{c}{Column scan [ms]} \\
  \cmidrule{2-5}
  & \texttt{DeepSqueeze} & \framework{} & \texttt{DeepSqueeze} & \framework{} \\
  \midrule
  \texttt{DASH}~\cite{dash} & 26.92 & 9.18 ~$\downarrow$& 1484.66 & 18.43~$\downarrow$ \\
  \texttt{Taxi trips}~\cite{taxi_2018} & 19.68 & 15.65~$\downarrow$ & 515.62 & 74.40~$\downarrow$ \\
  \texttt{Weekly Counts}[...]~\cite{weekly_counts_of_death} & 3.50 & 1.20~$\downarrow$ & 663.37 & 8.95~$\downarrow$ \\
  \bottomrule
\end{tabular}
\vspace{0.2cm}
\caption{Comparison to \texttt{DeepSqueeze}~\cite{deepsqueeze} on table size and virtualized column scan latency}
\label{tab:comparison}
\end{table}
\section{Discussion \& Future Work}\label{sec:conclusion}

\framework{} provides high space savings over Apache Parquet while having a negligible scan overhead. We believe that \emph{arbitrary} sparse functions, such as ``tiny'' neural networks, which ensure that the number of reference columns is minimized, can improve the current compression factors further. Moreover, scan performance could be accelerated by introducing additional metadata, such as min-max bounds, for the virtualized columns in the storage layer itself.

\bibliographystyle{unsrtnat} 
\bibliography{virtual}

\end{document}